\documentclass[aps,preprint,amsmath,amssymb]{revtex4-1}
\usepackage{slashed}
\usepackage{graphicx}
\usepackage{color}
\newcommand{\nn}{\nonumber}
\newcommand{\bd}{\begin{document}}
\newcommand{\ed}{\end{document}}
\newcommand{\bc}{\begin{center}}
\newcommand{\ec}{\end{center}}
\newcommand{\be}{\begin{eqnarray}}
\newcommand{\ee}{\end{eqnarray}}
\newcommand{\ba}{\begin{array}}
\newcommand{\ea}{\ed{array}}
\newcommand{\strich}[1]{#1  \! \! \slash}
\newcommand{\eqn}{\global\def\theequation}
\newcommand{\sw}{sin^2 \theta_W}
\newcommand{\fbd}{f_B}
\renewcommand{\thefootnote}{\alph{footnote}}
\newcommand{\se}{\section}
\newcommand{\sse}{\subsection}
\newcommand{\bi}{\bibitem}
\def\figcap{\section*{Figure Captions\markboth
     {FIGURECAPTIONS}{FIGURECAPTIONS}}\list
     {Figure \arabic{enumi}:\hfill}{\settowidth\labelwidth{Figure 999:}
     \leftmargin\labelwidth
     \advance\leftmargin\labelsep\usecounter{enumi}}}
\let\endfigcap\endlist \relax
\def\reflist{\section*{References\markboth
     {REFLIST}{REFLIST}}\list
     {[\arabic{enumi}]\hfill}{\settowidth\labelwidth{[999]}
     \leftmargin\labelwidth
     \advance\leftmargin\labelsep\usecounter{enumi}}}
\let\endreflist\endlist \relax

\def\Journal#1#2#3#4{{#1} {{\bf #2},} {#4} {(#3)}}
\def\NCA{Nuovo Cimento}
\def\NIM{Nucl. Instrum. Methods}
\def\NIMA{{Nucl. Instrum. Methods} A}
\def\NP{{Nucl. Phys.} }
\def\NPB{{Nucl. Phys.} B }
\def\NPA{{Nucl. Phys. A}}
\def\PLB{{Phys. Lett.}  B}
\def\PL{{Phys. Lett.}}
\def\PPSA{{Proc. Phys. Soc.} A}
\def\PRP{{ Phys. Rep.}}
\def\PRL{ Phys. Rev. Lett.}
\def\PR{{Phys. Rev.}}
\def\PRD{{Phys. Rev.} D}
\def\PRC{{Phys. Rev.} C}
\def\ZP{{Z. Phys.}}
\def\ZPC{{Z. Phys. C}}
\def\EPJ{{Eur. Phys. J.}}
\def\EPJC{{Eur. Phys. J.} C}
\def\ZPA{{Z. Phys.} A}
\def\MPL{{Mod. Phys. Lett.}}
\def\MPLA{{Mod. Phys. Lett.} A}
\def\CPC{Comput. Phys. Commun.}
\def\JHEP{{J. High Energy Phys.}}
\def\JPG{{J. Phys. G.}}
\def\SJNP{Sov. J. Nucl. Phys.}
\def\NCA{ Nuovo Cimento}
\def\NIM{ Nucl. Instrum. Methods}
\def\NIMA{{ Nucl. Instrum. Methods} A}
\def\NP{{ Nucl. Phys.}}
\def\ANP{{Adv. Nucl. Phys.}}
\def\CPC{{Comput. Phys. Commun.}}

\begin{document}
\title{Semileptonic decays of $\Lambda^{+}_{c}$ in light-front quark model with nonvalence contributions}

\author{Chong-Chung Lih\footnote{cclih123@gmail.com} and Chao-Qiang Geng\footnote{cqgeng@ucas.ac.cn}}
\affiliation{
Synergetic Innovation Center for Quantum Effects and Applications,
Hunan Normal University, Changsha 410081, China\\
School of Fundamental Physics and Mathematical Sciences,
Hangzhou Institute for Advanced Study, UCAS, Hangzhou 310024, China}

\date{\today}

\begin{abstract}
We investigate the exclusive semilpetonic decays of
$\Lambda^{+}_{c}\to (\Lambda/n) \ell^{+} \nu_{\ell}~(\ell=e,\mu)$  within the standard model by
using the  light-front quark model (LFQM).
The form factor behaviors are obtained from the effective treatment of
nonvalence contributions in addition to the valence ones
in the Drell-Yan-West frame due to the Bethe-Salpeter formalism.
Based on these
form factors, we find  that the decay
branching ratios of $\Lambda^{+}_{c}\to (\Lambda e^{+} \nu_{e},\,\Lambda \mu^{+} \nu_{e}
,\, n e^{+} \nu_{e},\, n \mu^{+} \nu_{e})$ are about $ (3.39,\,3.21,\,0.36,\,0.35)\%$
 with
 the non-valence contributions,
which are consistent
with  the recent experimental measurements at BESIII.
Furthermore, we  use the experimental data to fit the
$\beta$ parameters in the baryonic distribution amplitudes under the LFQM,
resulting in
 ($\beta_{\Lambda_{c}},\beta_{\Lambda}, \beta_{n})= (0.58\pm0.08, 0.52\pm0.08,0.44\pm0.04)$.

\end{abstract}

\maketitle

\se{Introduction}
The experimental and theoretical studies of
semileptonic decays in both baryons and mesons have played
a crucial role in building up the standard model (SM)
in particle physics.
Since 2015, the BESIII Collaboration has reported the
measurements  of the absolute branching fractions of the $\Lambda_{c}$ semilepton decays, i.e.,
$B(\Lambda^{+}_{c}\to \Lambda e^{+} \nu_{e}) = (3.63 \pm 0.38\pm 0.20)\%$
and $B(\Lambda^{+}_{c}\to \Lambda \mu^{+} \nu_{\mu}) = (3.49\pm 0.46\pm 0.27)\%$~\cite{exp1,exp2,exp3}.
Recently, the result of
$B(\Lambda^{+}_{c}\to n e^{+} \nu_{\mu})=(0.357\pm0.034\pm0.014)\%$
has  also been published~\cite{exp4}.
Currently, there have been many theoretical studies related to these decays,
such as those of the Lattice QCD~\cite{LQCD,LQCD2}, $SU(3)_F$ flavor symmetry~\cite{SU3},
QCD sum rule~\cite{QCDSR}, quark model (QM)~\cite{QM1,QM2,QM3},
Light-cone sum rule~\cite{LCSR2} and Light-front (LF) approach~\cite{LFQM1,LFQM2,LFQM3}.
The results of these semileptonic decays are useful for understanding the dynamics
in the charmed baryon physics.

Phenomenologically, the LF formalism offers a simple,
non-perturbative, and relativistic framework for calculating the hadronic form factors.
Since such a framework has been successfully applied to mesonic decay processes~\cite{lf2,lf3,lf4,lf5,lf6},
 it is natural to be used in the calculation of the form factors in the byronic decays~\cite{lf7,lf8,lf9,lf10,lf11}.
The aim of this study is to evaluate the hadronic form factors and determine the decay rates
of  $\Lambda^{+}_{c}\to (\Lambda/n) \ell \nu_{\ell}$ within the SM by
using the LFQM based on the LF quantization.
The LFQM in the present work employs a diquark picture~\cite{diquark,diquark2}, where
the baryon contains a heavy quark along with a light diquark.
This approximation makes the work similar to the meson case and
reduce the computational effort considerably. The non-perturbative interaction
between the two light quarks in the diquark picture
can be effectively absorbed into the diquark mass.
This helps us to analyze computationally the hadronic form factors in the entire (physical)
time-like regime and obtain the decay rates of the semileptonic decays of $\Lambda^{+}_{c}$.

The Drell-Yan-West  framework with $q^{+}=0$ is crucial  because
only the valence state contribution is required,
where $q^\mu$ is the momentum transfer and $q^{+}=q_{0}+q_{z}$ corresponds to its LF momentum
component.
This yields a form factor expressed by the overlap of the
hadronic LF wave functions before and after the decay
in the space-like ($q^{2} < 0$) region.
Some successful phenomenological calculations of the form factors
in the space-like region in the LFQM can be found in Refs.~\cite{lf7,lf10,jaus,LFg}.
However, some of the form factors in the exclusive processes may
receive higher Fock state contributions, i.e. zero modes in the $q^{+} = 0$
frame, within the framework of the LF quantization.
Therefore, it is necessary to take into account
the zero mode contributions in the space-like region.
Another option is to use the same ``$+$" component
but work in the range of $q^{+}\neq0$, that is, in the time-like ($q^{2} > 0$) region.
Physically, particle decay processes occur in a time-like zone, so
in principle,  form factors can be calculated using the $q^{+}\neq 0$
framework in the time-like region.
Similarly, those working in the time-like regime
may also receive contributions from non-valence Fock states,
which require higher Fock states of the wave function, as well as valence states~\cite{BS1}.
That is, we will inevitably encounter the non-valence diagram, so-called ``Z diagram",
generated by the production of quark-antiquark pairs.
Fortunately, effective methods for dealing with non-valence state contributions
in meson exclusion processes have been available in the literature~\cite{BS1,BS2,BS3}.
The goal of our new treatment is to map this procedure directly
onto the baryon semileptonic decay
and to extract the value of the parameter ``$\beta$" of the LFQM
by comparison with the experimental data.

This paper is organized as follows.
In Sec. II, we present the framework for the baryonic semileptonic decays.
Our numerical results and discussions are given in  Sec. III. We conclude in Sec. IV.

\section{A framework for semilepton decays of ${\cal B}_{i} \to {\cal B}_{f}\,\ell\,\nu_\ell$}

To study the exclusive spin-1/2  ${\cal B}_{i}$  to  ${\cal B}_{f}$ semileotonic decays, we start
with the charged current process of $Q \to q \ell^{+}\nu_{\ell}$ at  quark level, in which
the effective Hamiltonian  is  given by
\be
H_{eff}(Q\to q\, \ell\nu_\ell)= {G_F\over \sqrt{2}}V_{Qq}\bar{q}
\gamma_{\mu}(1-\gamma_5)Q\,\bar{\nu}_{\ell}\gamma_{\mu}(1-\gamma_5)\ell\,.
\label{he1}
\ee
Form Eq.(\ref{he1}), we get
\be
H_{eff}({\cal B}_{i} \to {\cal B}_{f}\,\ell\,\nu_\ell)= {G_F\over \sqrt{2}}V_{Qq}
\langle{\cal B}_{f}(P^{\prime},S^{\prime}=
\frac{1}{2})|\bar{q}
\gamma_{\mu}(1-\gamma_5)Q|{\cal B}_{i}(P,S=\frac{1}{2})\rangle
\,\bar{\nu}_{\ell}\gamma_{\mu}(1-\gamma_5)\ell\,,~
\label{he2}
\ee
with the transition matrix elements defined as follows:
\be
&&\langle{\cal B}_{f}(P^{\prime},S^{\prime}=
\frac{1}{2},S_{z}^{\prime})|\bar{q}\gamma^{\mu}
(1-\gamma_{5})Q|{\cal B}_{i}(P,S=\frac{1}{2},S_{z})\rangle \nonumber\\
&& =  \bar{u}_{\alpha}(P^{\prime},S_{z}^{\prime})
\Big[\gamma^{\mu} f_{1}(q^{2})
+i \frac{f_{2}(q^{2})}{M_{{\cal B}_{i}}}\sigma_{\mu\nu}q^{\nu}
+\frac{f_{3}(q^{2})}{M_{{\cal B}_{i}}} q_{\mu}\Big]
u(P,S_{z})\nonumber\\
& &\quad-\bar{u}_{\alpha}(P^{\prime},S_{z}^{\prime})
\Big[\gamma^{\mu} g_{1}(q^{2})
+i \frac{g_{2}(q^{2})}{M_{{\cal B}_{i}}}\sigma_{\mu\nu}q^{\nu}
+\frac{g_{3}(q^{2})}{M_{{\cal B}_{i}}} q_{\mu}\Big]\gamma_{5}u(P,S_{z}),
\label{transitionVA}
\ee
where $q^{\mu}=P^{\mu}-P^{\prime\mu}$ and $f_i(g_i)~(i=1,2,3)$ are the baryonic form factors.

The matrix elements in Eq.~(\ref{transitionVA}) can be applied to
the helicity amplitudes,
given by~\cite{lf11,HA1,HA2}
\begin{eqnarray}
H^{V(A)}_{\lambda_{{\cal B}_{f}} \lambda_{W}}
\equiv\langle {\bf {\cal B}_{f}}|(\bar q Q)_{V(A)}|{{\cal B}_{i}}\rangle
\varepsilon^\mu_W\,,
\label{helicityA}
\end{eqnarray}
where $\varepsilon^\mu_W$ is the polarization of the W boson,
and $\lambda_{{\cal B}_{f}({\cal B}_{i})}=\pm 1/2$
represent the helicity states of the final (initial) spin-1/2 baryons.
Based on the helicity conservation,
$\lambda_{{\cal B}_{i}}=\lambda_{{\cal B}_{f}}-\lambda_{W}$ is held.

The helicity amplitudes are related to the form factors through the
following expressions:
\begin{align}
H_{\frac{1}{2},t}^{V} & =-i\frac{\sqrt{Q_{-}}}{\sqrt{q^{2}}}
\left((M-M^{\prime})f_{1}+\frac{q^{2}}{M}f_{3}\right),\nonumber \\
H_{\frac{1}{2},0}^{V} & =-i\frac{\sqrt{Q_{-}}}{\sqrt{q^{2}}}
\left((M+M^{\prime})f_{1}-\frac{q^{2}}{M}f_{2}\right),\nonumber \\
H_{\frac{1}{2},1}^{V} & =i\sqrt{2Q_{-}}\left(-f_{1}
+\frac{M+M^{\prime}}{M}f_{2}\right),\nonumber \\
H_{\frac{1}{2},t}^{A} & =-i\frac{\sqrt{Q_{-}}}{\sqrt{q^{2}}}
\left((M+M^{\prime})f_{1}-\frac{q^{2}}{M}g_{3}\right),\nonumber \\
H_{\frac{1}{2},0}^{A} & =-i\frac{\sqrt{Q_{+}}}{\sqrt{q^{2}}}
\left((M-M^{\prime})g_{1}+\frac{q^{2}}{M}g_{2}\right),\nonumber \\
H_{\frac{1}{2},1}^{A} & =i\sqrt{2Q_{+}}\left(-g_{1}-\frac{M-M^{\prime}}{M}g_{2}\right).
\end{align}
where the subscript ``$t$" denote the helicity amplitude $H_{\frac{1}{2},t}^{V(A)}$
from the temporal component of the current of $(\bar q Q)_{V(A)}$.
The negative helicity amplitudes are defined by
\begin{equation}
H_{-\lambda^{\prime},-\lambda_{W}}^{V}=H_{\lambda^{\prime},
\lambda_{W}}^{V}\quad\text{and}\quad H_{-\lambda^{\prime},-\lambda_{W}}^{A}
=-H_{\lambda^{\prime},\lambda_{W}}^{A},
\end{equation}
while the helicity ones for the left-handed current are obtained as
\begin{equation}
H_{\lambda^{\prime},\lambda_{W}}=H_{\lambda^{\prime},
\lambda_{W}}^{V}-H_{\lambda^{\prime},\lambda_{W}}^{A}.
\end{equation}
Consequently,
the differential decay width of the semi-leptonic process reads
\begin{align}
\frac{d\Gamma_{L}}{dq^{2}}&=&\frac{G_{F}^{2}|V_{CKM}|^{2}}{(2\pi)^{3}}
\frac{(q^{2}-m_{\ell}^{2})^{2}\,\rm p_{cm}}{24M^{2} q^{2}}
\bigg\{\left(1+\frac{m_{\ell}^{2}}{2 q^{2}}\right)\left[|H_{\frac{1}{2},0}|^{2}+|H_{-\frac{1}{2},0}|^{2}
+|H_{\frac{1}{2},1}|^{2}+|H_{-\frac{1}{2},-1}|^{2}\right]\nonumber \\
&&+\frac{3 m_{\ell}^{2}}{2 q^{2}}\left(|H_{\frac{1}{2},t}|^{2}
+|H_{-\frac{1}{2},t}|^{2}\right)\bigg\}\,,
\label{Dwidth}
\end{align}
where $q^{2}$ is the lepton pair invariant mass,
$\rm p_{cm}$$ =\sqrt{Q_{+}Q_{-}}/2M$, $Q_{\pm}=(M\pm M^{\prime})^{2}-q^{2}$,
and $M$ ($M^{\prime}$) is the mass of the parent (daughter) baryon.
It is worth noting that the last term of Eq.~(\ref{Dwidth})
gives a small contribution to the decay rate
as it is associated with the lepton mass. Consequently, we may ignore the
calculations of $f_{3}(g_{3})$ in this study.

In order to calculate the form factors in the LFQM,
we treat the baryon as a bound state composed of three quarks $q_1$, $q_2$ and $q_3$,
where $q_{2,3}$ are combined into a diquark, expressed as $q_{[2,3]}$.
Explicitly, the baryon bound state with the total momentum
$P$ and spin $S$ can be written as~\cite{diquark2}
\be
|{\cal B}(P,S,S_{z})\rangle & = & \int\{d^{3}p_{1}\}
\{d^{3}p_{2}\}2(2\pi)^{3}\delta^{3}(\tilde{P}-\tilde{p}_{1}-\tilde{p}_{2})\nonumber \\
&  & \times\sum_{\lambda_{1},\lambda_{2}}\Psi^{SS_{z}}
(\tilde{p}_{1},\tilde{p}_{2},\lambda_{1},\lambda_{2})|q_{1}(p_{1},\lambda_{1})[q_2, q_3]
(p_{2},\lambda_{2})\rangle\,,
\label{boundstate}
\ee
where $q_{1}=c$ or $s$ denotes the active quark corresponding to
$\Lambda_c$ or $\Lambda$ baryon, $[q_2, q_3]$ represents the diquark,
$\Psi^{SS_{z}}$ corresponds to the momentum-space wave function and
$p_{1,2}$  are the on-mass-shell
LF momenta, and
\be
        \tilde p=(p^+, p_\bot)~, \quad p_\bot = (p^1, p^2)~,
                \quad p^- = {m^2+p_\bot^2\over p^+},
\ee with \be
        && p^+_1=x_1 P^+, \quad p^+_2=x_2 P^+, \quad x_1+x_2=1\,,\nn \\
        && p_{1\bot}=x_1 P_\bot+k_\bot, \quad p_{2\bot}=x_2
        P_\bot-k_\bot\,.
\label{Pfraction}
\ee
In Eq.~(\ref{Pfraction}), $(x,k_\perp)$ are the light-front relative momentum
variables, and $\vec{k}_\perp$ is the component of the internal
momentum $\vec{k}=(\vec{k}_\perp,k_z)$.

By the Melosh transformation~\cite{Melosh:1974cu},
it is more convenient to work with the following
representation of the wave function
\be
\Psi^{SS_{z}}(\tilde{p}_{1},\tilde{p}_{2},\lambda_{1},\lambda_{2})=
\frac{1}{\sqrt{2(p_{1}\cdot\bar{P}+m_{1}M_{0})}}\bar{u}(p_{1},\lambda_{1})
\Gamma_{l,m} u(\bar{P},S_{z})\phi(x,k_{\perp})\,,
\label{1/2}
\ee
where $\Gamma_{l,m}$ is the coupling vertex function of the decaying quark $q_{1}$
and the diquark in the baryon state.  For the  scalar diquark, the coupling vertex is $\Gamma_{s}=1$.
If the axial-vector diquark is involved, the vertex should be
\be
\Gamma_{A} & =&-\frac{1}{\sqrt{3}}\gamma_{5} \strich\epsilon^{*}(p_{2},\lambda_{2})\,.
\label{vertex_1/2}
\ee

The wave function of $\phi(x,k_{\perp})$ in Eq.~(\ref{1/2})
describes the momentum distribution of the constituent quarks in
the bound state. In this work, we use the Gaussian-type function, given as
\be
\phi(x,k_{\perp})=4\left(\frac{\pi}{\beta^{2}}\right)^{3/4}\sqrt{\frac{dk_{z}}{dx}}\exp
\left(\frac{-\vec{k}^{2}}{2\beta^{2}}\right)\,,
\label{DA}
\ee
where $\beta$ is the baryon shape parameter and
$k_z$ is defined by
\be
1-x=\frac{e_1-k_z}{e_1+e_2}\,,\quad  x=\frac{e_2+k_z}{e_1+e_2}\,,
\ee
with $e_i=\sqrt{m^2_{i}+\vec{k}^2}$. Subsequently, we have
\be
M_0=e_1+e_2\,, \quad
k_z=\frac{xM_0}{2}-\frac{m^2_{2}+k^2_{\perp}}{2xM_0}\,,
\ee
and
\be
M_0^2&=&{ m_{1}^2+k_\bot^2\over 1-x}+{ m_{2}^2+k_\bot^2\over  x}\,.
\ee

\begin{figure}[h]
\includegraphics{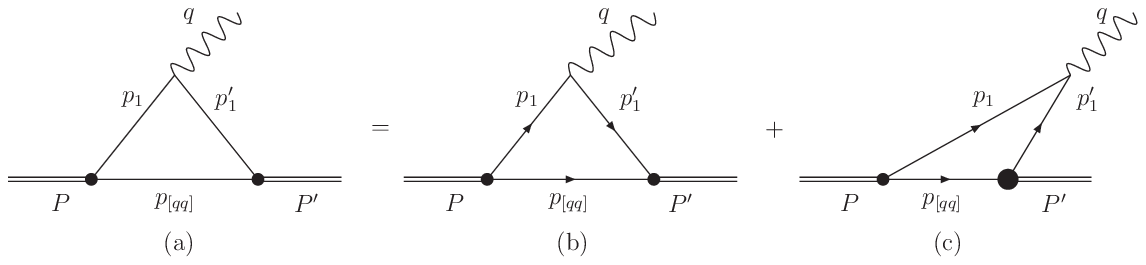}
\vskip 4cm
\caption{The effective treatment of the LF amplitude (a) can be displayed into
the LF valence part (b) in $0 < x <\alpha$ and the nonvalence one (c)
in $\alpha < x < 1$, where
the small and big dots of the mediator-quark vertices in (b) and (c) represent the LF ordinary
and nonvalence wavefunction vertices, respectively.}
\end{figure}

In our calculation, we obtain
\be
\alpha\equiv{ P'^{+}\over P^{+}}
=\frac{M^{2}_{B_{i}}+M^{2}_{B_{f}}-q^{2}+\sqrt{(M^{2}_{B_{i}}+M^{2}_{B_{f}}-q^{2})^{2}
-4 M^{2}_{B_{i}} M^{2}_{B_{f}}} }{2 M^{2}_{B_{i}}}\,.
\label{eq21}
\ee
Using the bound states of
$|{\cal B}_{i}(P,S,S_z)\rangle$ and $|{\cal B}_{f}(P^{\prime},S^{\prime},S_z^{\prime})\rangle$
in Eq.~(\ref{boundstate}) and the above identities,
we derive the matrix elements of the baryonic transition
in the LF frame. By considering the $\mu=+$ component,
the transition matrix elements are given by
\be
&&\langle {\cal B}_{f}(P^{\prime},S^{\prime},S_z^{\prime})|\bar q\gamma^\mu
(1-\gamma_5)Q|{\cal B}_{i}(P,S,S_z)\rangle\nonumber \\
&=&
N_{fs}\int{\{d^{3}p_{2}\}}
\frac{\Psi_{i}(x,{\bf k}_{\perp}) I^{+}
\Psi_{f}(x',{\bf k'}_{\perp})}{(1-x)(1-x')}\,,
\label{matrix}
\ee
where $I^{+}=\sum_{\lambda_{2}}\bar{u}(\bar{P}',S'_{z})
\left[\bar{\Gamma}^{\prime}_{S(A)}(\strich p_{1}^{\prime}+m_{1}^{\prime})
\gamma^{+}(1-\gamma_{5})(\strich p_{1}+m_{1})\Gamma_{S(A)}\right]u(\bar{P},S_{z})$,
$\bar \Gamma=\gamma^0 \Gamma^\dagger\gamma^0$,
and
the flavor spin factor $N_{fs}$ is the overlap factor of particle transformations,
which are given by different processes.
The trace term $I^{+}$ in Eq.~(\ref{matrix}) can be written as
the sum of the valence $I^{+}_{V}$ and non-valence
 $I^{+}_{NV}$ parts, as shown in Fig. 1a.
In the region  of $0<x<\alpha$ with $0<p^{+}< P'^{+}$ and $p^{-}_{[qq]}=p^{-}_{[qq]on}=
(m_{[qq]}^{2}+k^{2}_{\perp})/p^{+}_{[qq]}$, in which the subscript ``$on$"
indicates on the mass shell, called the valence region as seen in Fig. 1b,
the effective contribution of the LF valence amplitude is given by
\be
&&{\cal M}_{val}
=\frac{N_{fs}}{16\pi^3}
\int^{\alpha}_{0}dx\int d^{2}{\bf k}_{\perp}
\frac{\Psi_{i}(x,{\bf k}_{\perp}) I^{+}_{V}
\Psi_{f}(x',{\bf k'}_{\perp})}{(1-x)(1-x')},\nonumber \\
&&I^{+}_{V}=\bar{u}(\bar{P}^{\prime},S_{z}^{\prime})\bar{\Gamma}^{\prime}
(\strich p_{1}^{\prime}+m_{1}^{\prime})\gamma^{+}(1-\gamma_{5})
(\strich p_{1}+m_{1})\Gamma u(\bar{P},S_{z})\,.
\label{ValM}
\ee
Following Refs.~\cite{LFcal1,LFcal2,LFcal3,LFcal4}, we obtain the transition form factors:
\be
f_{1_{V}}(q^{2})&=&\frac{N_{fs}}{16\pi^3}
\int^{\alpha}_{0}dx\int d^{2}{\bf k}_{\perp}
\frac{\Psi_{i}(x,{\bf k}_{\perp}) \Psi_{f}(x',{\bf k'}_{\perp})}
{8(1-x)(1-x')P^{+}P^{\prime+}}\nonumber \\
&&\times{\rm Tr} [ (\slashed P+M_{0})\gamma^{+}(\slashed P^{\prime}+M_{0}^{\prime})
(\slashed p_{1}^{\prime}+m_{1}^{\prime})\gamma^{+}
(\slashed p_{1}+m_{1}) ] \nonumber \\
&=&\frac{N_{fs}}{16\pi^3}
\int^{\alpha}_{0}dx\int d^{2}{\bf k}_{\perp}
\frac{\Psi_{i}(x,{\bf k}_{\perp}) \Psi_{f}(x',{\bf k'}_{\perp})}
{(1-x)(1-x')}\nonumber \\
&&\times[k_{\perp}\cdot k_{\perp}^{\prime}+(x_{1}M_{0}+m_{1})
(x_{1}^{\prime}M_{0}^{\prime}+m_{1}^{\prime})]\nonumber \\
g_{1_{V}}(q^{2})&=& \frac{N_{fs}}{16\pi^3}
\int^{\alpha}_{0}dx\int d^{2}{\bf k}_{\perp}
\frac{\Psi_{i}(x,{\bf k}_{\perp}) \Psi_{f}(x',{\bf k'}_{\perp})}
{8(1-x)(1-x')P^{+}P^{\prime+}}\nonumber \\
&&\times{\rm Tr} [ (\slashed P+M_{0})\gamma^{+}\gamma_{5}(\slashed P^{\prime}+M_{0}^{\prime})
(\slashed p_{1}^{\prime}+m_{1}^{\prime})\gamma^{+}\gamma_{5}
(\slashed p_{1}+m_{1}) ]\nonumber \\
&=&\frac{N_{fs}}{16\pi^3}
\int^{\alpha}_{0}dx\int d^{2}{\bf k}_{\perp}
\frac{\Psi_{i}(x,{\bf k}_{\perp}) \Psi_{f}(x',{\bf k'}_{\perp})}
{(1-x)(1-x')}\nonumber \\
&&\times[-k_{\perp}\cdot k_{\perp}^{\prime}+(x_{1}M_{0}+m_{1})
(x_{1}^{\prime}M_{0}^{\prime}+m_{1}^{\prime})]\nonumber \\
\frac{f_{2_{V}}(q^{2})}{M}&=& \frac{N_{fs}}{16\pi^3}
\int^{\alpha}_{0}dx\int d^{2}{\bf k}_{\perp}
\frac{\Psi_{i}(x,{\bf k}_{\perp})
\Psi_{f}(x',{\bf k'}_{\perp})}{8(1-x)(1-x')P^{+}P^{\prime+}q_{\perp}^{\nu}}\nonumber \\
&&\times{\rm Tr} [ (\slashed P+M_{0})\sigma^{\nu +}(\slashed P^{\prime}+M_{0}^{\prime})
(\slashed p_{1}^{\prime}+m_{1}^{\prime})\gamma^{+}
(\slashed p_{1}+m_{1}) ] \nonumber \\
&=&\frac{N_{fs}}{16\pi^3}
\int^{\alpha}_{0}dx\int d^{2}{\bf k}_{\perp}
\frac{\Psi_{i}(x,{\bf k}_{\perp}) \Psi_{f}(x',{\bf k'}_{\perp})}
{(1-x)(1-x')q_{\perp}^{2}}\nonumber \\
&&\times[-(m_{1}+x_{1}M_{0})k_{\perp}^{\prime}\cdot q_{\perp}
+(m_{1}^{\prime}+x_{1}^{\prime}M_{0}^{\prime})k_{\perp}\cdot q_{\perp}]\nonumber \\
\frac{g_{2_{V}}(q^{2})}{M}&=& \frac{N_{fs}}{16\pi^3}
\int^{\alpha}_{0}dx\int d^{2}{\bf k}_{\perp}
\frac{\Psi_{i}(x,{\bf k}_{\perp})
\Psi_{f}(x',{\bf k'}_{\perp})}{8(1-x)(1-x')P^{+}P^{\prime+}q_{\perp}^{\nu}}\nonumber \\
&&\times{\rm Tr} [ (\slashed P+M_{0})\sigma^{\nu +}\gamma_{5}(\slashed P^{\prime}+M_{0}^{\prime})
(\slashed p_{1}^{\prime}+m_{1}^{\prime})\gamma^{+}\gamma_{5}
(\slashed p_{1}+m_{1}) ]\,\nonumber \\
&=&\frac{N_{fs}}{16\pi^3}
\int^{\alpha}_{0}dx\int d^{2}{\bf k}_{\perp}
\frac{\Psi_{i}(x,{\bf k}_{\perp}) \Psi_{f}(x',{\bf k'}_{\perp})}
{(1-x)(1-x')q_{\perp}^{2}}\nonumber \\
&&\times[-(m_{1}+x_{1}M_{0})k_{\perp}^{\prime}\cdot q_{\perp}
-(m_{1}^{\prime}+x_{1}^{\prime}M_{0}^{\prime})k_{\perp}\cdot q_{\perp}]\,.
\label{ffscalar}
\ee
where $\nu = 1, 2$ and the variables of $(x',k'_\perp)$ are the light-front relative momentum
variables of ${\cal B}_{f}(P^{\prime},S^{\prime},S_z^{\prime})$
with the definitions given by replacing $x\to x'$ in Eq.~(\ref{Pfraction}).
In this manuscript, we will not calculate $f_3(g_3)$ for the reasons explained in the
calculations of helicity amplitudes early.

In the nonvalence region of $\alpha<x<1$, $P'^{+}<p^{+}<P^{+}$ and $p_{1}^{-}=p^{-}_{1on}=
(m_{1}^{2}+k^{2}_{1\perp})/p^{+}_{1}$,
as shown in Fig. 1c,
the trace term in Eq.~(\ref{matrix}) can be separated into
the on-shell propagating and instantaneous parts of $I^{\mu}_{on}$ and $I^{\mu}_{inst}$ via
\be
\slashed p+m=(\slashed p_{on}+m)+\frac{1}{2}\gamma^{+}(p^{-}-p^{-}_{on})\,,
\label{eq24}
\ee
respectively. The effective contribution can be found as
\be
{\cal M}_{non-val}&=&\frac{N_{fs}}{16\pi^3}
\int^{1}_{\alpha}dx
\int d^{2}{\bf k}_{\perp}\frac{\Gamma_g(x,{\bf k}_{\perp})I^{+}_{NV}}
{(1-x)(x'-1)}\Psi_i(x,{\bf k}_{\perp})\nonumber\\
&\times&
\int \frac{dy}{y(1-y)}\int d^2{\bf l}_{\perp}
{\cal K}(x,{\bf k}_{\perp};y,{\bf l}_{\perp})
\Psi_{f}(y,{\bf l}_{\perp})\,.
\label{NVmatrix}
\ee
where $I^{+}_{NV}$ is the trace term in the non-valence region.
Substituting Eq.~(\ref{eq24}) into $I^{+}$, we can obtain
$I^{+}_{NV}=I^{+}_{V}(p_{i}^{-}=p^{-}_{ion}=(m_{i}^{2}+k^{2}_{i\perp})/p^{+}_{i})
+I^{+}_{inst}$.
The LF vertex function of a gauge boson
$\Gamma_g$ in Eq.~(\ref{NVmatrix}) corresponds to the LF
energy denominator with its explicit form  given by~\cite{BS1,BS3,LFg}
\be
\Gamma_g^{-1}(x,{\bf k}_\perp)=
\alpha\biggl[\frac{q^2}{1-\alpha} -
\biggl(\frac{{\bf k}^2_\perp + m^2_1}{1-x}
+\frac{{\bf k'}^2_\perp + m'^{2}_{1}}{\alpha-x}\biggr)
\biggr].
\label{gaugeWF}
\ee
The trace terms in Eqs.~(\ref{ValM}) and (\ref{NVmatrix}),
both corresponding to the products of the initial and final LF spin wave functions,
can be obtained by off-shell Meloche transformations.
The form factors related to the non-valence diagram $I^+_{NV}$ are given by
\be
f_{1_{NV}}(q^{2})&=&\frac{N_{fs}}{16\pi^3}
\int^{1}_{\alpha}dx
\int d^{2}{\bf k}_{\perp}\Gamma_g(x,{\bf k}_{\perp})\Psi_i(x,{\bf k}_{\perp}) \nonumber \\
&\times&\frac{[k_{\perp}\cdot k_{\perp}^{\prime}+((1-x_{1})M_{0}+m_{1})
((1-x_{1}^{\prime})M^{\prime}+m_{1}^{\prime})]+I^{+}_{inst}}
{(1-x)(x'-1)}\nonumber\\
&\times&
\int \frac{dy}{y(1-y)}\int d^2{\bf l}_{\perp}
{\cal K}(x,{\bf k}_{\perp};y,{\bf l}_{\perp})
\Psi_{f}(y,{\bf l}_{\perp})\,,
\nonumber
\label{f1NV}
\ee
\be
g_{1_{NV}}(q^{2})&=& \frac{N_{fs}}{16\pi^3}
\int^{1}_{\alpha}dx
\int d^{2}{\bf k}_{\perp}\Gamma_g(x,{\bf k}_{\perp})\Psi_i(x,{\bf k}_{\perp})\nonumber\\
&\times&\frac{[-k_{\perp}\cdot k_{\perp}^{\prime}+((1-x_{1})M_{0}+m_{1})
((1-x_{1}^{\prime})M^{\prime}+m_{1}^{\prime})]+I^{+}_{inst}}
{(1-x)(x'-1)}\nonumber\\
&\times&
\int \frac{dy}{y(1-y)}\int d^2{\bf l}_{\perp}
{\cal K}(x,{\bf k}_{\perp};y,{\bf l}_{\perp})
\Psi_{f}(y,{\bf l}_{\perp})\,,
\nonumber
\label{g1NV}
\ee
\be
\frac{f_{2_{NV}}(q^{2})}{M}&=& \frac{N_{fs}}{16\pi^3}
\int^{1}_{\alpha}dx
\int d^{2}{\bf k}_{\perp}\Gamma_g(x,{\bf k}_{\perp})\Psi_i(x,{\bf k}_{\perp})\nonumber\\
&\times&\frac{
[(m_{1}+(1-x_{1})M_{0})k_{\perp}^{\prime}\cdot q_{\perp}
-(m_{1}^{\prime}+(1-x_{1}^{\prime})M^{\prime})k_{\perp}\cdot q_{\perp}]+I^{\prime +}_{inst}}
{(1-x)(x'-1)}\nonumber\\
&\times&
\int \frac{dy}{y(1-y)}\int d^2{\bf l}_{\perp}
{\cal K}(x,{\bf k}_{\perp};y,{\bf l}_{\perp})
\Psi_{f}(y,{\bf l}_{\perp})\,,
\nonumber
\label{f2NV}
\ee
\be
\frac{g_{2_{NV}}(q^{2})}{M}&=& \frac{N_{fs}}{16\pi^3}
\int^{1}_{\alpha}dx
\int d^{2}{\bf k}_{\perp}\Gamma_g(x,{\bf k}_{\perp})\Psi_i(x,{\bf k}_{\perp})\nonumber\\
&\times&\frac{
[-(m_{1}+(1-x_{1})M_{0})k_{\perp}^{\prime}\cdot q_{\perp}
-(m_{1}^{\prime}+(1-x_{1}^{\prime})M^{\prime})k_{\perp}\cdot q_{\perp}]+I^{\prime +}_{inst}}
{(1-x)(x'-1)}\nonumber\\
&\times&
\int \frac{dy}{y(1-y)}\int d^2{\bf l}_{\perp}
{\cal K}(x,{\bf k}_{\perp};y,{\bf l}_{\perp})
\Psi_{f}(y,{\bf l}_{\perp})\,.
\label{g2NV}
\ee
where $I^{+}_{inst}=(1-x)(M^{2}-M_{0}^{2})(m'_1+(1-x')M')$ and
$I^{\prime +}_{inst}=-(1-x)(M^{2}-M_{0}^{2})k^{\prime}_{\perp}\cdot q_{\perp}$.
The complete form factors are $f(g)_{j}=f(g)_{j_{V}}+f(g)_{j_{NV}}$ ($j=1,2$).
It is worth noting that the instantaneous contribution exists
in the non-valence diagram only when the ``$+$" component is used.
The vertex of the nonvalence wave function is usually obtainable
from the Bethe-Salpeter (BS) amplitude in the BS theory~\cite{FVNV,BS1}.
The LF BS equation is expressed as~\cite{BS1,BSEQ2,BSEQ3}
\be
(M^{2}-M^{2}_{0})\Psi'_{f}(x_{i},{k}_{i\perp})
=\int [dy][d^2{\bf l}_{\perp}]
{\cal K}(x_{i},{\bf k}_{i\perp};y,{\bf l}_{\perp})
\Psi_{f}(y,{\bf l}_{\perp})\,.
\label{LFBS}
\ee
Both valence and nonvalence BS amplitudes can be regarded as solutions of Eq.~(\ref{LFBS}).
The normal and nonvalence BS amplitudes correspond  to  $x > \alpha$ and  $x < \alpha$, respectively.
In Fig. 1c, the nonvalence B-S amplitude
is the analytic continuation of the valence BS amplitude.
In the LFQM, the relationship between the BS amplitudes of
two regions is given in Refs.~\cite{BS1,BS2,BS3}.
However, for the integral equation of Eq.~(\ref{LFBS})
it is necessary to use  the nonperturbative QCD method to obtain the kernel.

\begin{figure}[h]
\includegraphics{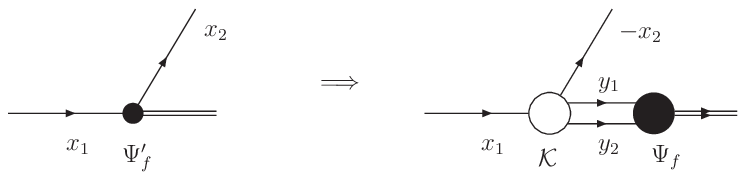}
\vskip 4cm
\caption{Nonvalence vertex (small black dot) linked to an ordinary
light-front wave function (big black dot) through a relevant operator ${\cal K}$ (big white dot).}
\label{NVvertex}
\end{figure}

The relevant operator ${\cal K}$ in Eq.~(\ref{LFBS}) is the BS core,
which in principle contains contributions from high Fock states.
It is the high Fock component of the bound state that is related
to the lowest Fock component with this kernel and describes the
connection of non-valence vertices (small black dots) to ordinary
light front wave functions (large black dots) through the correlation operator
${\cal K}$ (large white dots), as shown in the Fig.~\ref{NVvertex}, where
$\Psi'(x_{i},\,k_{i\perp})$ represents the non-valent B-S amplitude.
We can imagine that the correlation operator ${\cal K}$ is a bridge
between the non-valent vertices and the ordinary light front wave function.
So the kernel ${\cal K}$ will be a function of all internal momenta (x; k; y; l).
We define that
$G_{{\cal B}_{i}{\cal B}_{f}}\equiv\int[dy][d^2{\bf l}_{\perp}]
{\cal K}(x_{i},{\bf k}_{i\perp};y,{\bf }_\perp)\Psi_f(y,{\bf l}_{\perp})$,
which depends only on $x$ and ${\bf k}_{\perp}$.
The range of the momentum fraction $x$ relies on the external momenta for the embedded states.

\se{Numerical Results And Discussions}

\sse{Form factors}
To numerically evaluate the exclusive transition form factors in the LFQM,
we choose to work directly in the time-like region.
Instead of assuming $q_{\perp}\neq 0$, we  first introduce
$q_{\perp}= 0 $  in the final numerical calculation process,
although we could also get the results in the time-like region by directly taking
$q^{2}_{\perp}\to -q^{2}$ for those in the space-like one.
In our calculation, we use~\cite{diquark2}
\be
m_{u,d} &=& 0.25\,,~~~~  m_{s} = 0.38\,,~~~~\,
m_{[qq']}=0.5\,,~~~~\, m_{c}=1.2~ {\rm GeV}.
\label{quarkmass}
\ee
where the quark masses are the constituent masses used in the quark model.
In Eq.~(\ref{NVmatrix}), $G_{{\cal B}_{i}{\cal B}_{f}}$  is treated  as
a constant in the range of $1.0 \sim 6.0$, which was previously tested in
some exclusively semileptonic mesonic decay processes and shown
to be a good approximation for processes with a small momentum transfer~\cite{BS1,BS2,BS3}.
Explicitly, in our numerical evaluation we choose
$G_{{\cal B}_{i}{\cal B}_{f}}=5(2)$ and the flavor-spin factors
$N_{fs}$ are  $\frac{1}{\sqrt{3}}$($\frac{1}{\sqrt{2}}$) for
$\Lambda_{c}\to \Lambda (n)$ transitions \cite{LFQM1}, respectively.

To describe the momentum $q^2$ behaviors,
we parameterize the form factors in the double-pole forms of
\be
F(q^2)=\frac{F(0)}{1+a(q^2/m_{P}^{2})+b(q^4/m_{P}^{4})}\,,
\ee
where $m_{P}=6.0$ GeV is the $M_{B}$ scale for the
modes of $\Lambda_{c}\to \Lambda (n)$,
and $(F(0),a,b)$ can be determined in the numerical analysis.
Our results for the form factors are given in Table \ref{Table1}.
In our results, the errors come from uncertainties in the shape parameters of $\beta$.

\begin{figure}[t!]
\includegraphics[width=3.1in]{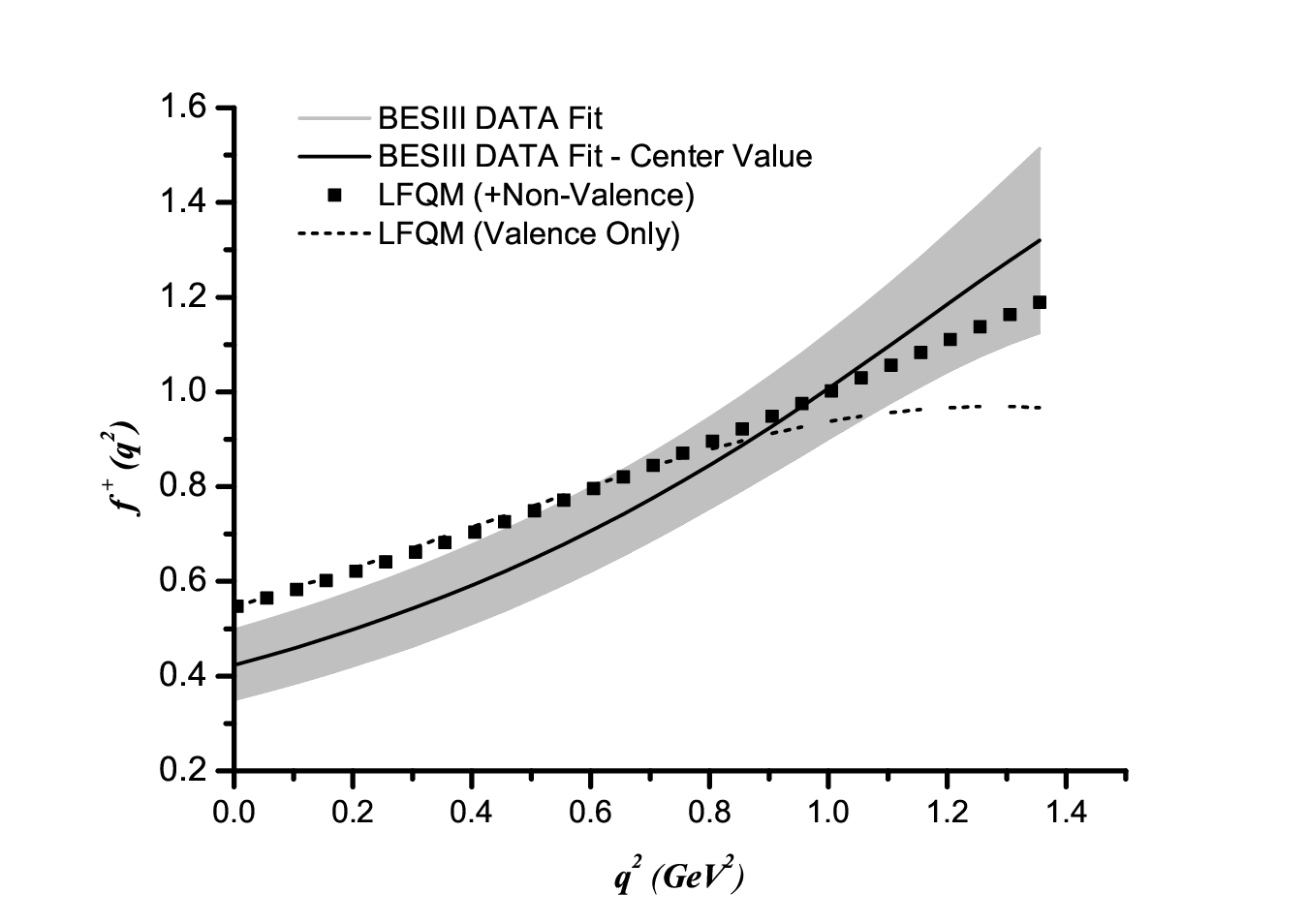}
\includegraphics[width=3.1in]{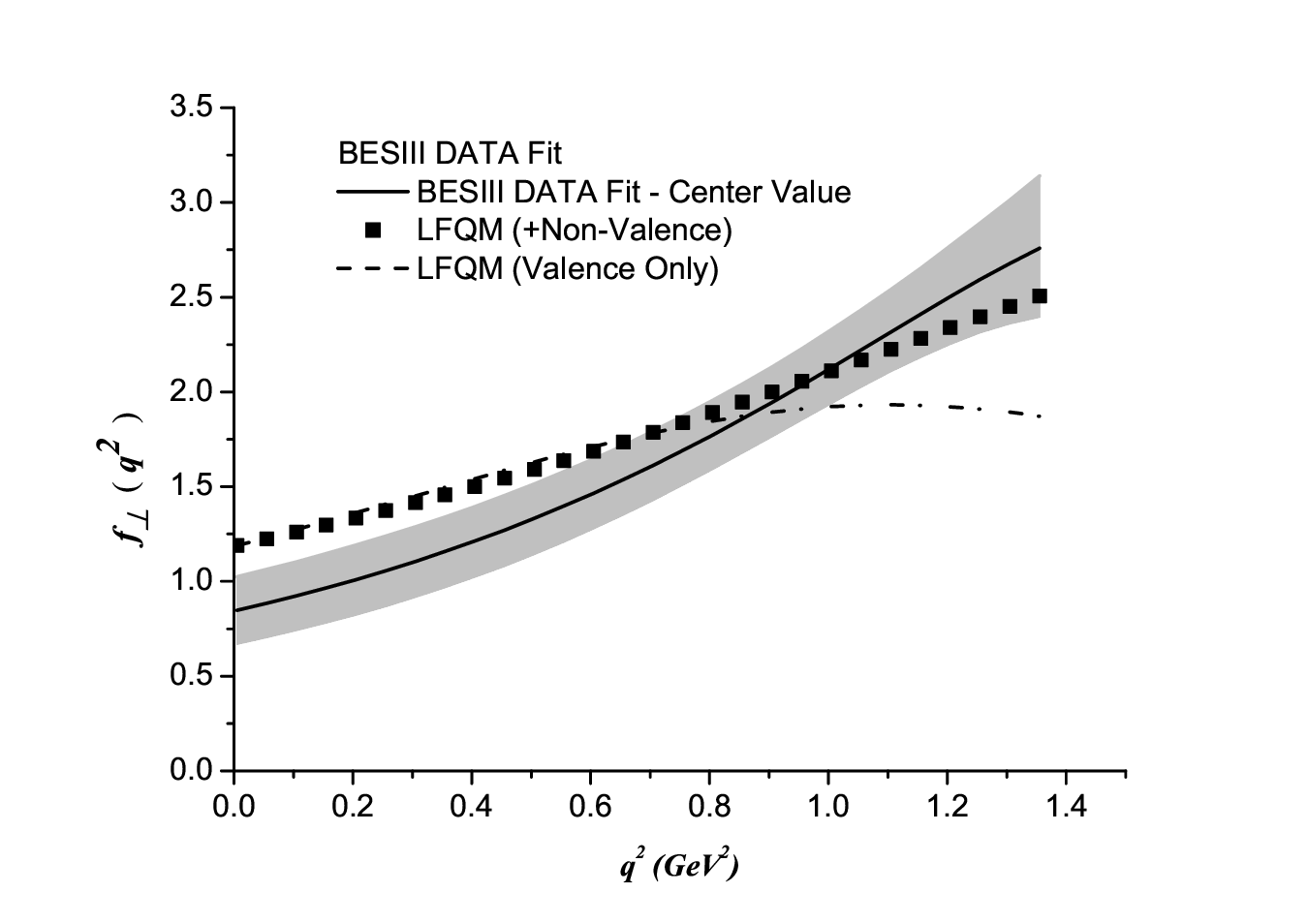}
\includegraphics[width=3.1in]{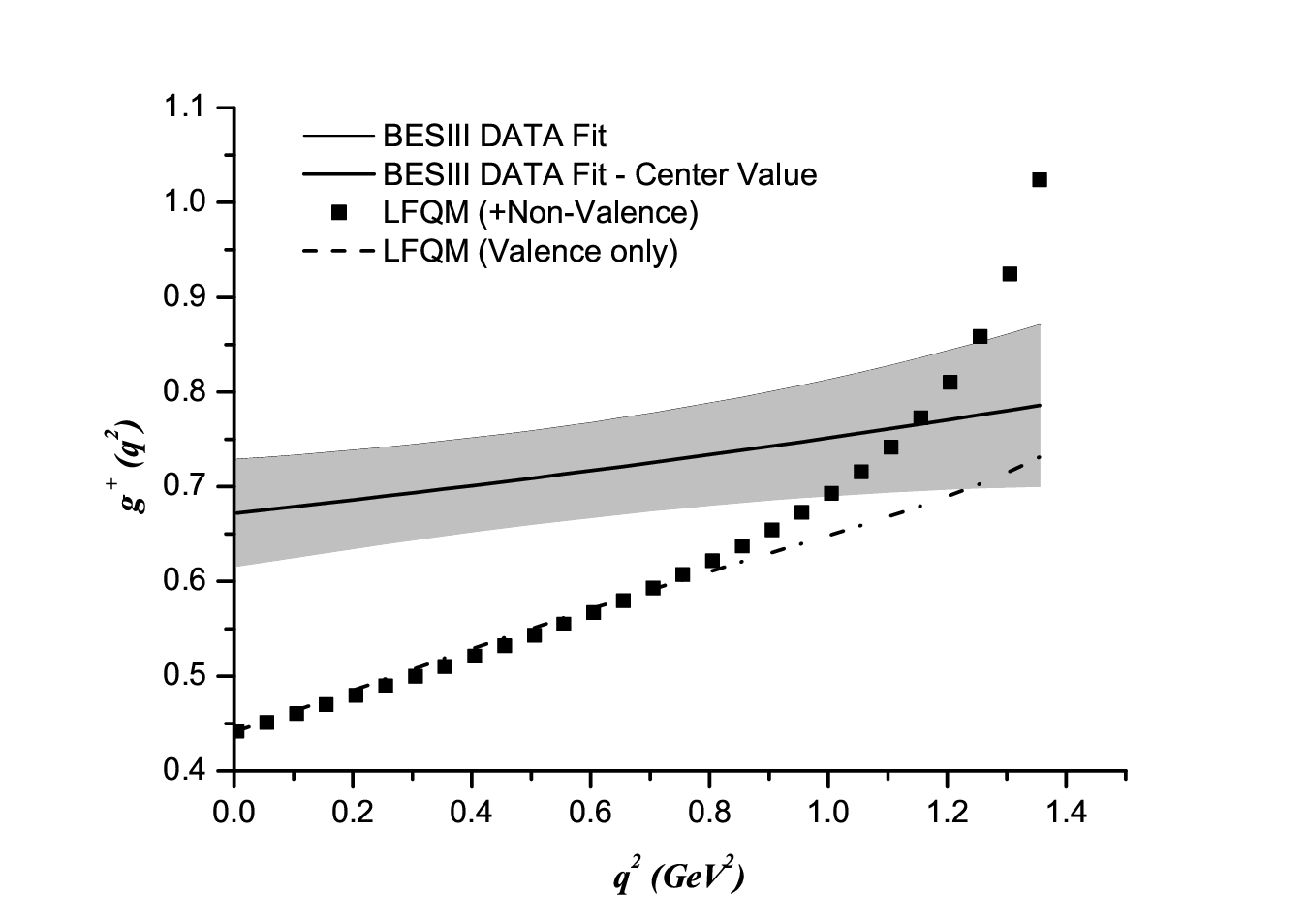}
\includegraphics[width=3.1in]{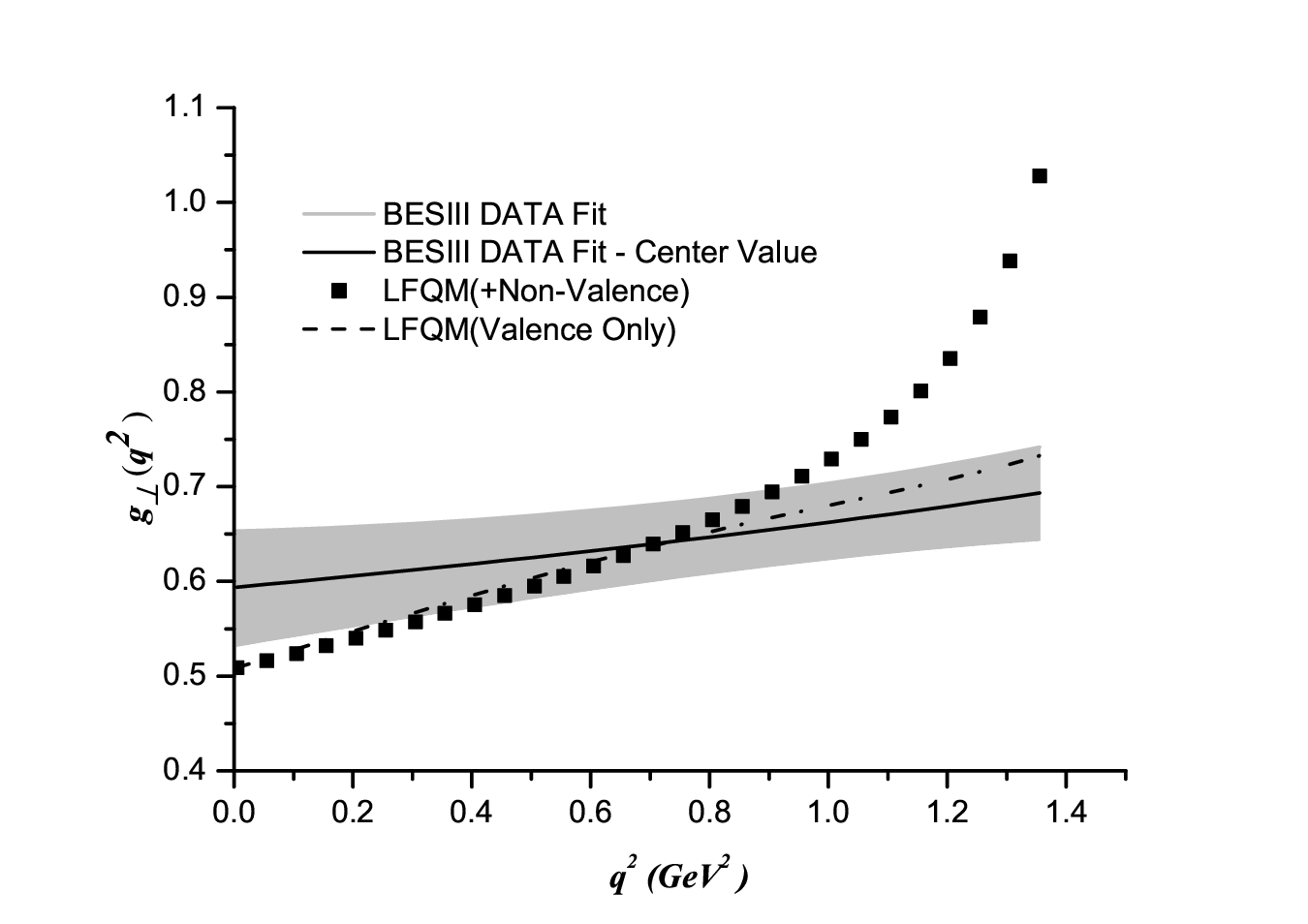}
\caption{Form factors of $\Lambda_{c} \to \Lambda$,
where the gray bands represent the uncertainties.}
\label{fig2}
\end{figure}

To find
 the value of $\beta$, we first
used the possible ranges of $\beta_{c[qq]} = 0.46\sim 0.76,~\beta_{s[qq]} = 0.4\sim 0.6$ in  Refs.~\cite{LFQM1,LFQM2}  as
the initial values to  calculate the decay branching ratios,
and then took the mean values of all branching ratios with
 one-standard-deviations as their errors.
 Consequently, we compare the experimental results in Fig.~\ref{fig2} with the
form factors corresponding to the mean values of the obtained branching ratios
to determine the most suitable combinations of $\beta$ values. 
Based on our results,  the $\beta$ values
 for  $\Lambda_{c}$, $\Lambda$ and $n$ in the LFQM are determined to  be
\be
\label{beta values}
\beta_{c[qq]}= 0.58\pm0.08,~~~~~~~\beta_{s[qq]}= 0.52\pm0.08,
~~~~~~~\beta_{u[qq]}= 0.40\pm0.04.
\ee
where $(\beta_{c[qq]}-\beta_{s[qq]})\sim 0.08$ and
$(\beta_{c[qq]}-\beta_{u[qq]})\sim 0.18$,
so that the form factors corresponding to the changes
in the $q^2$ spectra clearly have better behaviors
to compare with the experimental results.
We note that the ratios of the uncertainties for the form factors in Table~\ref{Table1}
are not the same due to the different values of $\beta_\Lambda$ and $\beta_n$ in the 
$\Lambda_c\to\Lambda$ and $\Lambda_c\to n$ transitions.
\begin{table}[htbp]
\caption{From factors of the $\Lambda_{c} \to (\Lambda$/n) transitions}
\vskip 0.2in
\label{Table1}
\begin{tabular}{|c||c|c|c|} \hline
$\Lambda_{c} \to \Lambda$ & $F(0)$ & $a$ & $b$
\\ \hline \hline
$f_{1}$& $0.546^{+0.030}_{-0.039}$
& $-19.02$ & $179.03$
\\ \hline
$f_{2}$ & $-0.43^{+0.039}_{-0.015}$
& $-22.07$ & $183.29$
\\ \hline
$g_{1}$ & $0.441^{+0.004}_{-0.009}$
& $-10.86$ & $59.74$
\\ \hline
$g_{2}$ & $0.131^{+0.016}_{-0.018}$
& $-7.84$ & $-372.00$
\\ \hline \hline
$\Lambda_{c} \to n$ & $F(0)$ & $a$ & $b$
\\ \hline \hline
$f_{1}$ & $0.595^{+0.033}_{-0.040} $
& $-16.14$ & $142.204$
\\ \hline
$f_{2}$ & $-0.478^{+0.032}_{-0.037} $
& $-15.76$ & $129.07$
\\ \hline
$g_{1}$ & $0.490^{+0.008}_{-0.025} $
& $-8.51$ & $23.22$
\\ \hline
$g_{2}$ & $0.057^{+0.0118}_{-0.0128} $
& $1.002$ & $-369.925$
\\ \hline
\end{tabular}
\end{table}
According to  Ref.~\cite{BESIII4}, we can rewrite the four form factors as
\be
f^{+}(q^{2}) & =&\frac{1}{M+M^{\prime}}
\left((M+M^{\prime})f_{1}-\frac{q^{2}}{M}f_{2}\right),\nonumber \\
f_{\perp}(q^{2}) & =&\left(f_{1}
-\frac{M+M^{\prime}}{M}f_{2}\right),\nonumber \\
g^{+}(q^{2}) & =&\frac{1}{M-M^{\prime}}
\left((M-M^{\prime})g_{1}+\frac{q^{2}}{M}g_{2}\right),\nonumber \\
g^{\perp}(q^{2}) & =&\left(g_{1}+\frac{M-M^{\prime}}{M}g_{2}\right)\,,
\label{HA3}
\ee
which  correspond to the four helicity amplitudes, respectively.
In Fig.~\ref{fig2}, we present our evaluations of the form factors as
functions of $q^2$ for the
$\Lambda_{c}\to \Lambda$ transition, where the BESIII data fits are also given.
Considering Eq.~(\ref{HA3}) and the results of BESIII in Fig.~\ref{fig2},
$f^{+}= f_{1}\sim 0.41$ at $q^{2}=0$.
Comparing to $f_{\perp}$ in  Fig.~\ref{fig2},
$f_{1}$ and $f_{2}$ must have opposite signs.
Our predictions are very close to those of the BESIII values
for $f^{+}$ and $f_{\perp}$ at $q^{2} = 0$,
but significantly smaller than the BESIII one for $g^{+}(0)$.
This leads to a branching ratio smaller than that of BESIII.
It is also worth mentioning that our calculations
as well as the LQCD one indicate that
$f^{+}(0) > g^{+}(0)$, but the opposite result is given from BESIII.

\begin{figure}[t!]
\includegraphics[width=3.1in]{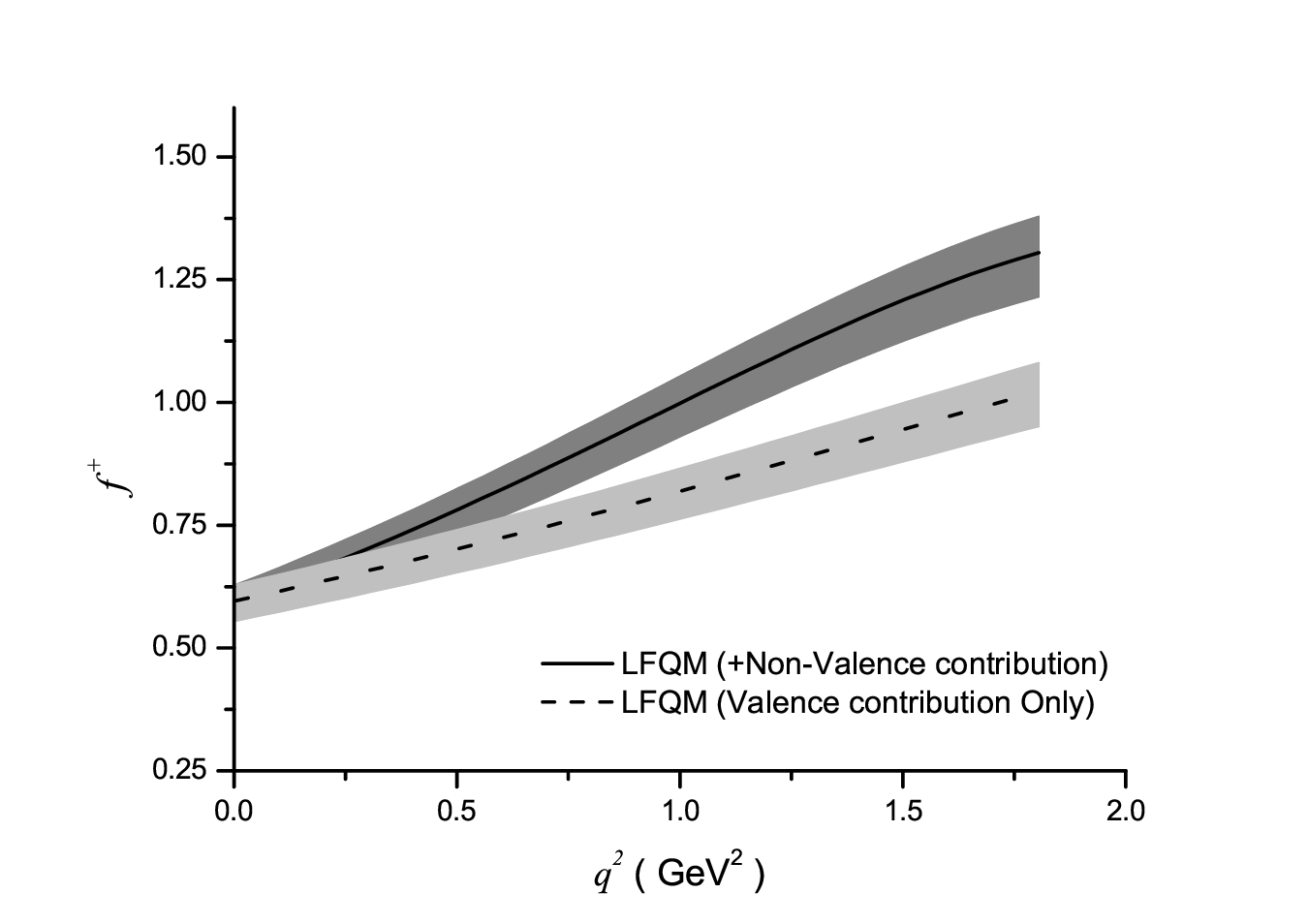}
\includegraphics[width=3.1in]{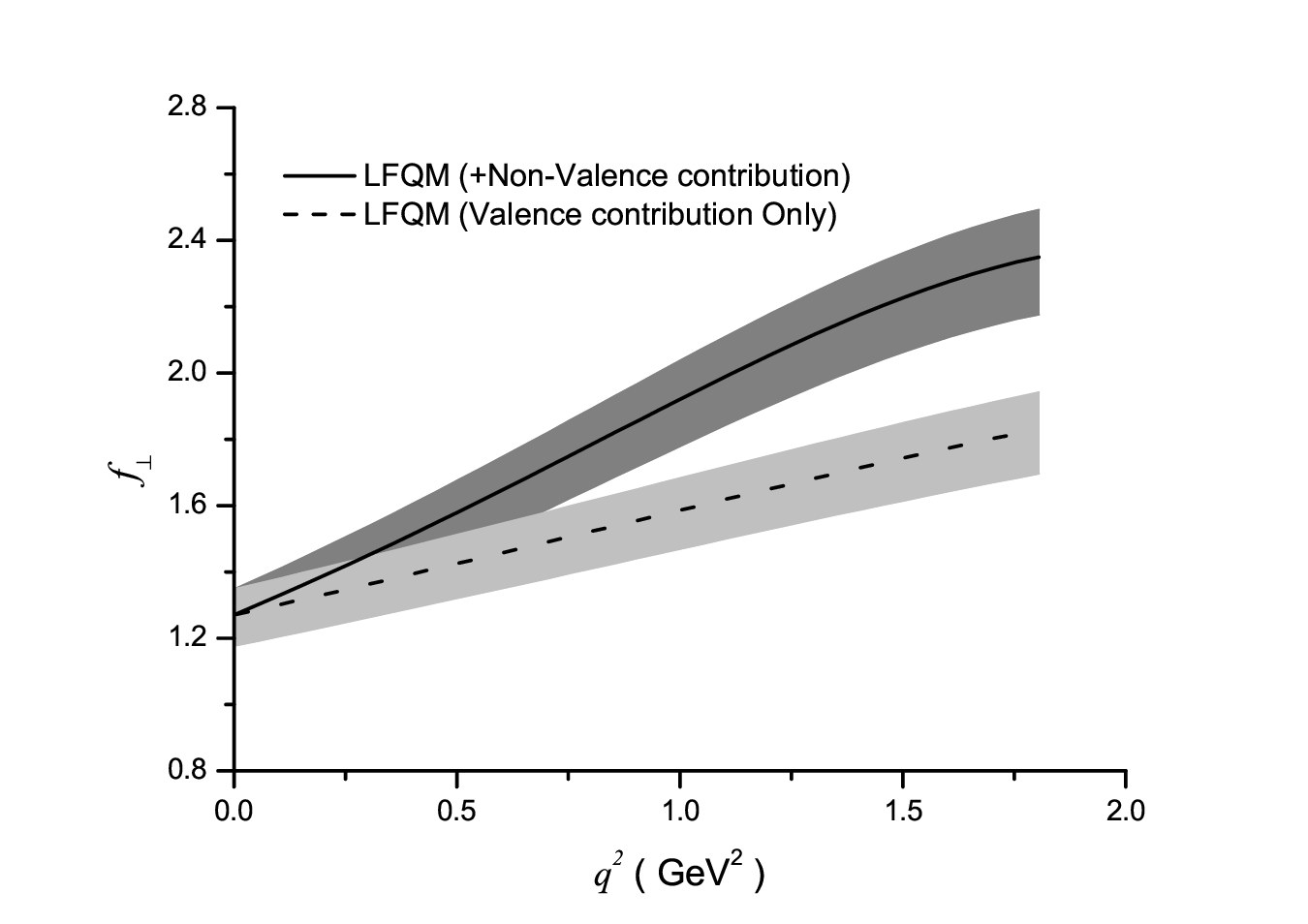}
\includegraphics[width=3.1in]{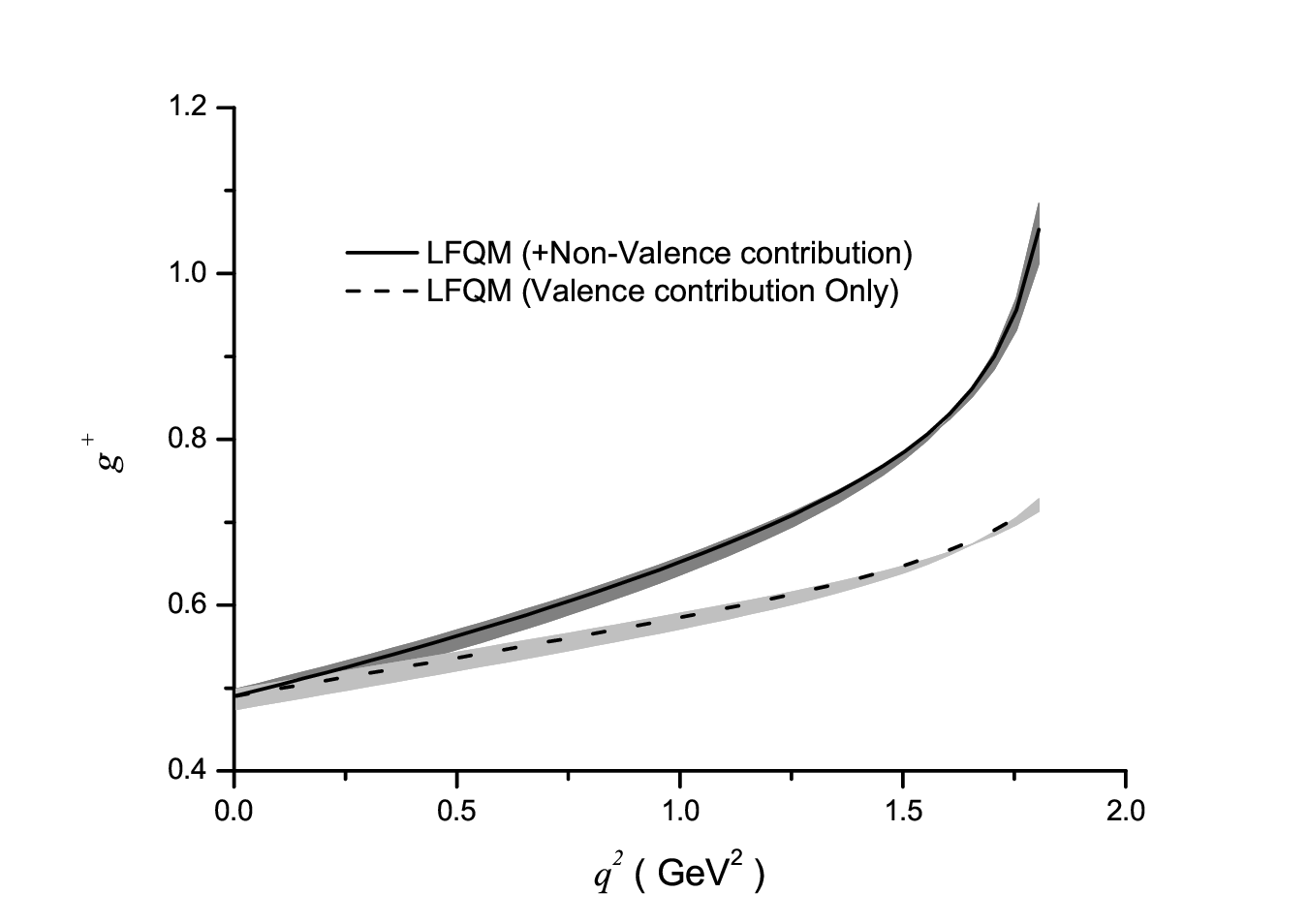}
\includegraphics[width=3.1in]{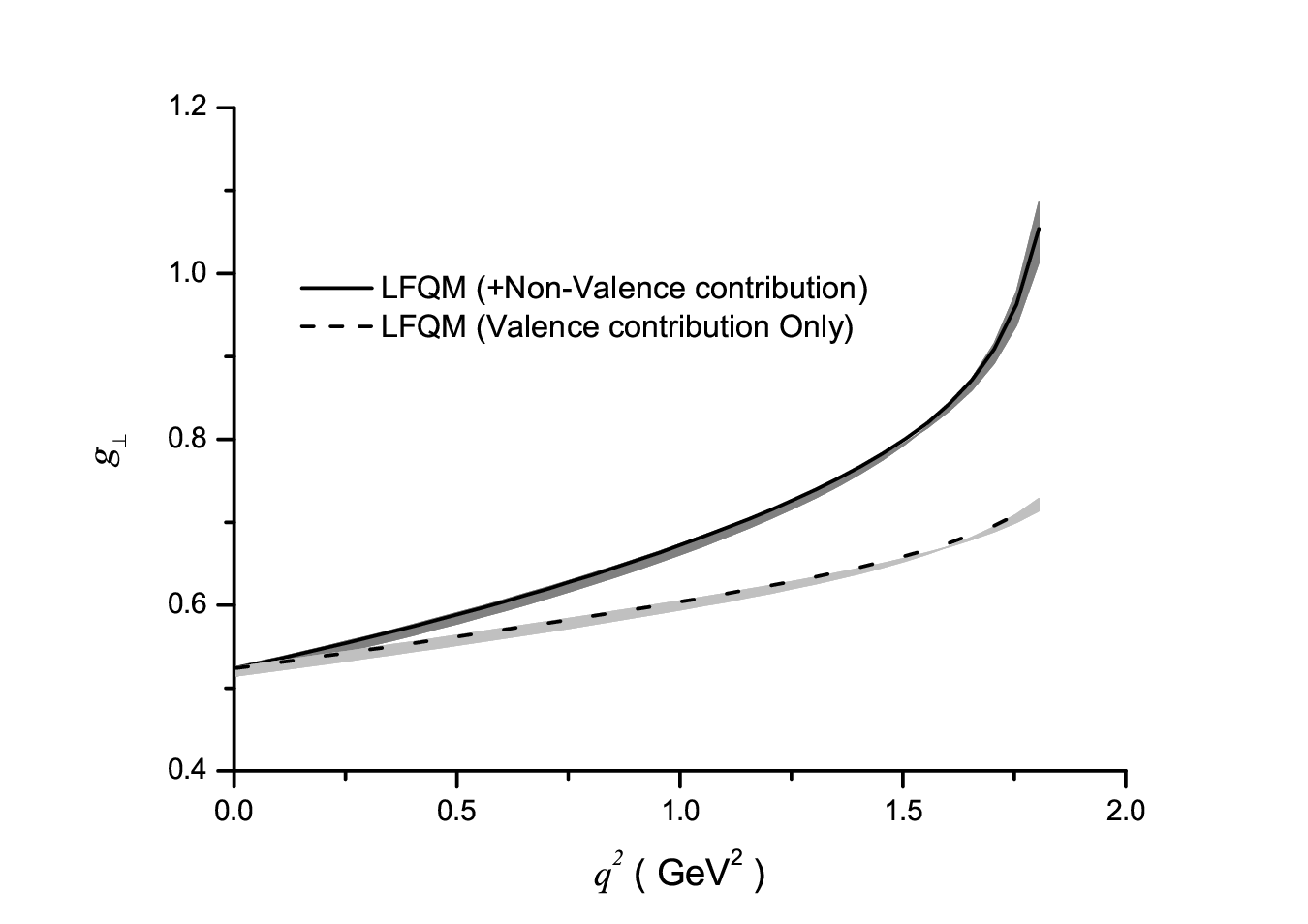}
\caption{Form factors of $\Lambda_{c} \to n$,
where the gray bands show the uncertainties.}
\label{figx}
\end{figure}

Fig.~\ref{fig2} shows that the four form factors exhibit similar
behaviors throughout the $q^{2}$ spectra. Among them,
$f^{+}$ and $f_{\perp}$ are more significant and closer to the BESIII values.
However, for $g^{+}$ and $g_{\perp}$, especially in the high $q^{2}$ region,
the LFQM calculation yields steeper slopes.
Obviously, our LFQM results with the non-valence contributions
are consistent with the BESIII data as well as those of the LQCD~\cite{BESIII4,LQCD}.
Additionally, Fig.~\ref{figx} displays the form factors
for the $\Lambda_{c}\to n$ transition,
similar to those for the $\Lambda_{c}\to \Lambda$ transition.

\sse{Decay branching ratios }

The amplitudes of $\Lambda_{c} \to (\Lambda/n)\,\ell\,\nu_\ell$ contain some
independent mixtures of helicity components,  described by $h^{V(A)}_{\lambda, \lambda_{p}}$, where
$\lambda$ and $\lambda_{p}$
represent the helicity components of the final baryon and W propagator, respectively.
From Eq.~(\ref{Dwidth}), we can easily separate the integrals of the
longitudinal and transverse polarization asymmetries.
The differential branching ratios  are given by
\be
{d\,\Gamma(\Lambda_{c} \to (\Lambda/n)\,\ell\,\nu_\ell)
\over{\Gamma(\Lambda_{c})\,dq^2}}&=&\frac{G_{F}^{2}\,V_{cq}^{2}}
{(2\,\pi)^{3}}\frac{(q^{2}-m_{\ell}^{2})^{2}\rm p_{cm}}{24M^{2} q^{2}\,\Gamma(\Lambda_{c})}\nonumber \\
&&\bigg\{\left(1+\frac{m_{\ell}^{2}}{2 q^{2}}\right)
\left[|H_{\frac{1}{2},0}|^{2}+|H_{-\frac{1}{2},0}|^{2}
+|H_{\frac{1}{2},1}|^{2}+|H_{-\frac{1}{2},-1}|^{2}\right]\nonumber \\
&&+\frac{3 m_{\ell}^{2}}{2 q^{2}}\left(|H_{\frac{1}{2},t}|^{2}
+|H_{-\frac{1}{2},t}|^{2}\right)\bigg\}\,.
\label{llr}
\ee
In Table \ref{Table2}, we show the decay branching ratios of
$\Lambda_{c} \to (\Lambda/n)\,\ell\,\nu_\ell~(\ell=e,\,\mu)$
in various LFQMs along with the BESIII data, where I and II represent our predictions with and without
nonvalence  contributions, respectively.
\begin{table}[htbp]
\caption{Decay branching ratios of $\Lambda_{c} \to (\Lambda/n)\,\ell\,\nu_\ell~(\ell=e,\,\mu)$.}
\vskip 0.2in
\label{Table2}
\begin{tabular}{|c||c|c||c|c|} \hline
 Result   & ${\cal B}(\Lambda_{c}\to\Lambda\,e\,\nu_e)$ & ${\cal B}(\Lambda_{c}\to\Lambda\,\mu\,\nu_\mu)$
&${\cal B}(\Lambda_{c}\to n\,e\,\nu_e)$ & ${\cal B}(\Lambda_{c}\to n\,\mu\,\nu_\mu)$
\\ \hline \hline
 I &$(3.39^{+0.26}_{-0.32})\% $ & $(3.214^{+0.245}_{-0.326})\% $
& $(0.361^{+0.02}_{-0.029})\% $ & $(0.345^{+0.019}_{-0.029})\% $
\\ \hline
 II & $(3.18^{+0.26}_{-0.32})\% $ & $(3.004^{+0.222}_{-0.298})\% $
& $(0.282^{+0.016}_{-0.023})\% $ & $(0.267^{+0.016}_{-0.021})\% $
\\ \hline
\cite{LFQM1} & $1.63\%$ &-& $0.201\%$ &-
\\ \hline
\cite{LFQM2} & $(3.36\pm0.87)\%$ & $(3.21\pm0.85)\%$ & $(0.36\pm0.15)\%$ &
$(0.34\pm0.14)\%$
\\ \hline
\cite{LFQM3} & $(4.04\pm0.75)\%$ & $(3.90\pm0.73)\%$ &-&-
\\ \hline\hline
BESIII~\cite{exp1,exp2,exp3} & $(3.63\pm0.38\pm0.20)\% $
& $(3.49\pm 0.46\pm 0.27)\% $ & $(0.357\pm0.034\pm0.014)\% $ & -
\\ \hline
\end{tabular}
\end{table}
It can be seen that the contribution with including
the non-valence part is about 10$\%$ more than that without it.
Note that the excess mainly comes from regions with higher $q^2$.
Compared with the other LFQM calculations, our results agree with
those in Ref.~\cite{LFQM2} well and  Ref.~\cite{LFQM3} within the allowed errors.
In particular, Ref.~[12], which works directly with three-quark state wave functions,
gives almost the same results as ours.
This suggests that using the diquark picture in the LFQM is a reasonable assumption.
However, the values in Ref.~\cite{LFQM1} are about only half of our ones.
Our predictions are also consistent with the BESIII data.
We note that the errors in our results are only
taken from the uncertainties in the $\beta$ factors.
Clearly, if the uncertainties of other parameters,
such as the quark masses and different shape parameters,
are added, these errors will increase.

\begin{figure}[t!]
\includegraphics[width=3.1in]{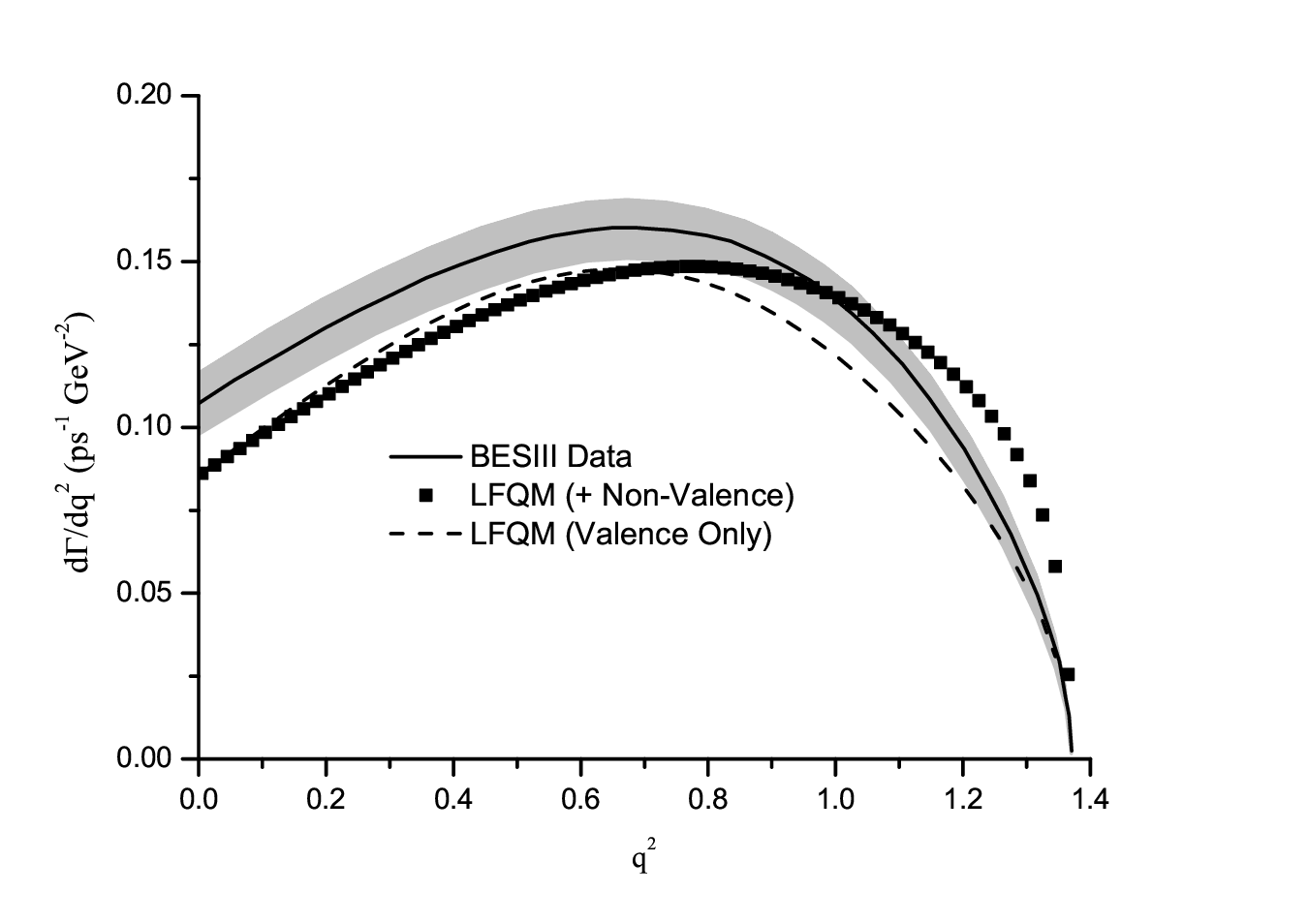}
\includegraphics[width=3.1in]{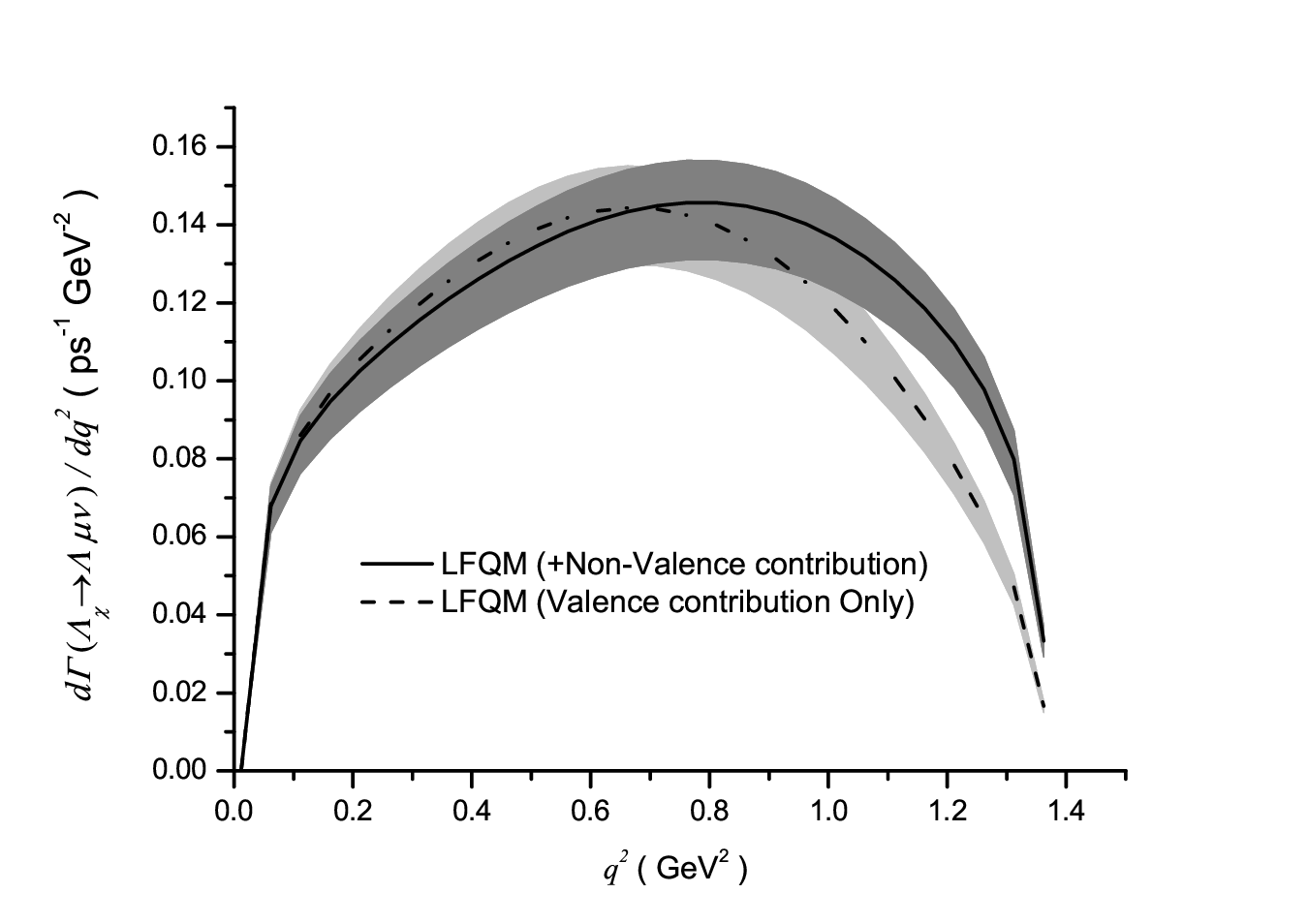}
\caption{Differential decay widths of $\Lambda_{c} \to \Lambda\,\ell \,\nu_\ell~(\ell=e,\,\mu)$, where
the gray bands show the total uncertainties.}
\label{fig3x}
\end{figure}

\begin{figure}[t!]
\includegraphics[width=3.1in]{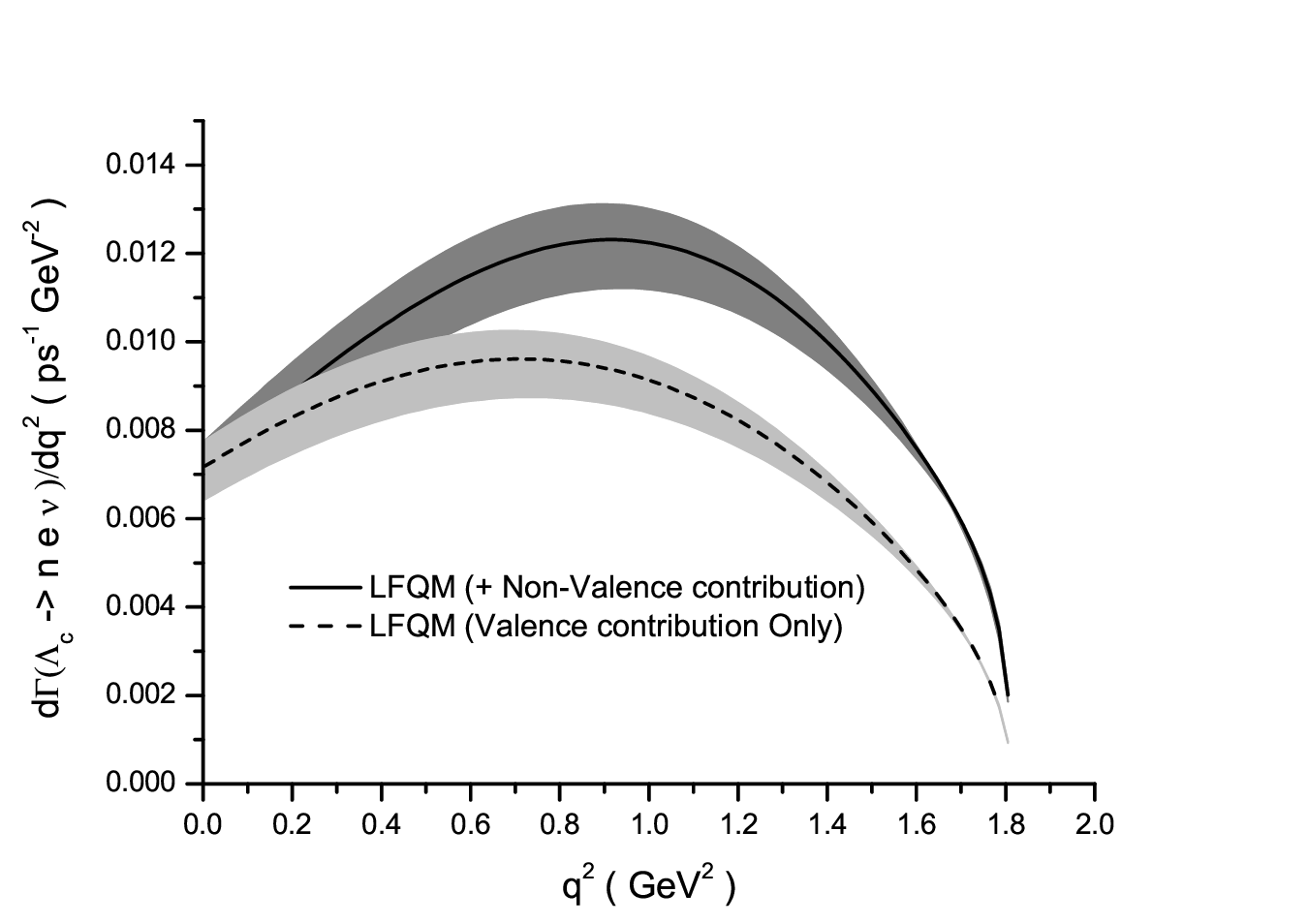}
\includegraphics[width=3.1in]{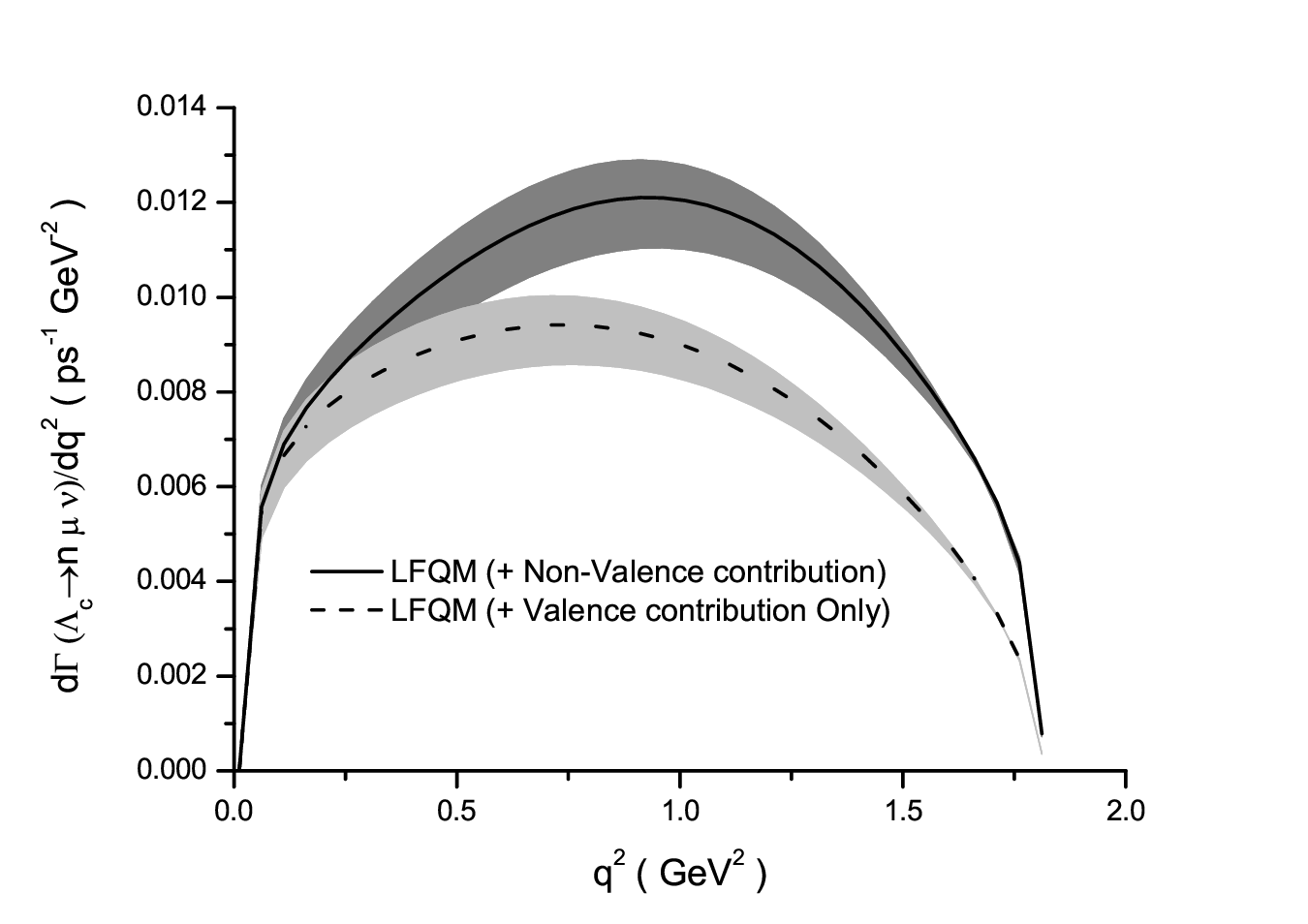}
\caption{Differential decay widths of 
$\Lambda_{c} \to n\,\ell \,\nu_\ell~(\ell=e,\,\mu)$,
where the gray bands show the total uncertainties.}
\label{fig4x}
\end{figure}

The differential decay rates of
$\Lambda_{c} \to (\Lambda,n)\,\ell \,\nu_\ell~(\ell=e,\,\mu)$ as functions of $q^2$
are displayed in Figs.~\ref{fig3x} and \ref{fig4x}, respectively.
As shown in Fig.~\ref{fig3x}, we see that a
considerable consistency across the $q^2$ region between our results and those from BESIII
for the differential decay rates.


Finally, we discuss the sensitivities of  the input parameters to our results.
 In Table II, the central values of our results for the decay branching ratios
 correspond to the central ones of $\beta_{c[qq]}$, $\beta_{s[qq]}$ and $\beta_{u[qq]}$ 
 in Eq.~(\ref{beta values}), while those of upper (lower) uncertainties are calculated by taking the 
  for the upper (lower) values of $\beta$, respectively.
Note that our results are based on fixed values of 
$m_q$, and $m_{[qq]}$ in Eq.~(\ref{quarkmass}).
Clearly, different choices of these parameters also affect the outcome.
For instance, for  $\Lambda^{+}_{c}\to \Lambda \ell^{+} \nu_{\ell}$,
if the diquark mass $m_{[qq]}$, which ranges from $0.4$ to $0.7$~GeV~\cite{diquark2}, 
increases $20\%$ 
without altering the baryonic $\beta$ values,
the decay branching ratio will also increase  about 5$\%$.
However, increasing the c-quark mass in $\Lambda_c$ by 0.2 GeV ($\sim$15$\%$) 
would decrease the branching ratio by about 15$\%$. As for the 
s-quark in $\Lambda$, the change of $m_s$ has a negligible effect on our results.

\se{Conclusion}

We have studied the exclusive semilepton
decays of  $\Lambda^{+}_{c}\to (\Lambda,n) \ell^{+} \nu_{\ell}$ 
in the SM by using the LFQM.
In our calculations, we have analyzed
the form factors of the baryonic transitions
based on the  BS formalism in $q^{+} > 0$,
efficiently dealing with nonvalence contributions.
The $q^2$ behaviors of the form factors are
consistent with the  BESIII data fits.
As shown in Table II, our results with including nonvalence contributions agree with the
existing experimental data of BESIII as well as  other theoretical evaluations.
In addition, we have also extracted the parameters of possible
$\beta$ values in the LFQM for $\Lambda_{c}$, $\Lambda$ and n, respectively.
Our phenomenological predictions for the semileptonic decays of $\Lambda_c$
can be tested by the experiments at various charmed baryon facilities.
Finally, applications to other similar baryonic 
decay processes will be considered elsewhere.

\section*{Acknowledgments}
 This work is supported in part by the National Key
Research and Development Program of China under Grant No. 2020YFC2201501
and  the National Natural Science Foundation of China (NSFC) under Grant No. 12347103.

\end{document}